\documentclass[pra,twocolumn,showkeys,showpacs,superscriptaddress,amsmath,amsfonts]{revtex4-1}
\usepackage[utf8x]{inputenc}
\usepackage{microtype}
\usepackage{lmodern}
\usepackage{graphicx}
\usepackage{hyperref}
\usepackage{color}
\usepackage{bm}

\newcommand{\bra}[1]{\langle#1\rvert}
\newcommand{\ket}[1]{\lvert#1\rangle}
\newcommand{\mean}[1]{\langle#1\rangle}
\newcommand{\vect}[1]{\bm{{#1}}}

\begin{document}

\title{Orbiton-magnon interplay in the spin-orbital polarons
       of KCuF$_3$ and LaMnO$_3$}

\author{Krzysztof Bieniasz}
\email{krzysztof.bieniasz@uj.edu.pl}
\affiliation{\mbox{Marian Smoluchowski Institute of Physics, Jagiellonian University,
             prof. S. {\L}ojasiewicza 11, PL-30348 Krak\'ow, Poland}}
\affiliation{Department of Physics and Astronomy, University of British Columbia,
             Vancouver, British Columbia, Canada V6T 1Z1}
\affiliation{\mbox{Quantum Matter Institute, University of British Columbia,
             Vancouver, British Columbia, Canada V6T 1Z4}}

\author{Mona Berciu}
\affiliation{Department of Physics and Astronomy, University of British Columbia,
             Vancouver, British Columbia, Canada V6T 1Z1}
\affiliation{\mbox{Quantum Matter Institute, University of British Columbia,
             Vancouver, British Columbia, Canada V6T 1Z4}}

\author{Andrzej M. Ole\'s}
\affiliation{\mbox{Marian Smoluchowski Institute of Physics, Jagiellonian University,
             prof. S. {\L}ojasiewicza 11, PL-30348 Krak\'ow, Poland}}
\affiliation{Max Planck Institute for Solid State Research,
             Heisenbergstra{\ss}e 1, D-70569 Stuttgart, Germany}

\date{\today}

\begin{abstract}
We present a quasi-analytical solution of a
spin-orbital model of KCuF$_{3}$, using the variational method for
Green's functions. By analyzing the spectra for different partial
bosonic compositions as well as the full solution, we show that
hole propagation needs both orbiton and magnon excitations to
develop, but the orbitons dominate the picture.
We further elucidate the role of the different bosons by analyzing
the self-energies for simplified models, establishing that because
of the nature of the spin-orbital ground state,
magnons alone do not produce a full quasiparticle band, in contrast
to orbitons. Finally, using the electron-hole transformation between
the $e_g$ states of KCuF$_3$ and LaMnO$_3$ we suggest the qualitative
scenario for photoemission experiments in LaMnO$_3$.
\end{abstract}


\maketitle

\section{Introduction}
\label{sec:intro}

In compounds with intraorbital Coulomb interaction $U$ electrons
localize and the ground state is determined by effective superexchange
interactions. The properties of such systems doped by holes may be very
different \cite{Ima98}. The hole propagation in a two-dimensional (2D)
antiferromagnetic (AF) square lattice~\cite{Lee06} is promoted by
quantum fluctuations and is controlled by the superexchange
$J$~\cite{Mar91}. In systems with orbital degrees of freedom the
superexchange is of spin-orbital form
\cite{Tok00,Kug82,Fei97,*FOZ98,Fei99,Kha00,Kha01,*Hor08,*Kha04,Ole05,Kha05,Jan09,
Woh11,*Woh13,*Che15,Ere11,Cor12,Nas13,*Nas15,Brz15,*Brz16,*Brz17}.
One finds then the whole plethora of different behaviors and the
details of hole propagation depend on the type of $3d$ orbitals
involved and on the system's dimension
\cite{vdB00,Bal01,Yin01,Ish05,Dag08,*Woh08,Woh09,*Mona,Wro10}.

Perhaps the most complex situation arises in the $e_g$ orbital model
where both the orbital superexchange \cite{vdB99} and the kinetic
energy, which does not conserve the orbital flavor \cite{Fei05}, are
radically different from those in the spin $t$-$J$ model \cite{Cha77}.
In a ferromagnetic (FM) compound, the orbital interactions are
Ising-like in a one-dimensional model~\cite{Dag04} but quantum
fluctuations increase via 2D towards three-dimensional (3D) orbital
model \cite{vdB99}.
This is in contrast to the SU(2) symmetric interactions in the AF
Heisenberg model \cite{Mar91}. However, when the ground state is AF and
spin excitations contribute as well, hole propagation is dominated by
the orbital excitations \cite{Woh09,*Mona} and holes are quasi-localized.
It is a challenging question to ask what happens when AF and alternating
orbital (AO) order appear in orthogonal directions. It was suggested
that \emph{a priori} only one type of excitations, either magnons or
orbitons, will control hole propagation in photoemission for LaMnO$_3$
\cite{Bal01}, but this was not verified until now.

The purpose of this paper is to study in a systematic
way the electron (hole) propagation in the 3D spin-orbital model for
KCuF$_3$ (LaMnO$_3$) at low temperature $T\to 0$. We construct the
$t$-$J$-like Hamiltonian with both spin and $e_g$ orbital degrees of
freedom and show that while orbitons dominate the picture, magnons are
also important to explain the low-energy quasiparticle (QP). Such a QP
state is a spin-orbital polaron, defined as a moving charge dressed by
both spin and orbital excitations \cite{Woh09,*Mona}. This polaron is
analogous to the spin polaron in the undoped cuprates
\cite{Mar91,Che00,Fle01,Wro08,Mie11}.
Surprisingly, in the present case magnons do not slow down the polaron
considerably or make it incoherent.

The remaining of the paper is organized as follows.
We introduce the spin-orbital model in Sec. \ref{sec:model}.
In Sec. \ref{sec:var} we describe the variational method used
to derive the spectral function from a systematic expansion in
terms of bosons standing for orbiton or magnon excitations.
Numerical results are presented and discussed in Sec. \ref{sec:res}.
The paper is summarized by the conclusions in Sec. \ref{sec:summa}.
We present also an Appendix with
the analytic results obtained for a one-boson expansion.

\section{The spin-orbital model}
\label{sec:model}

We first discuss KCuF$_{3}$ which is conceptually easier,
being a tetragonal system (cubic in the first approximation), with
Cu($d^{9}$) ions placed in octahedral cages of fluorides. Their crystal
field splits the $3d$ orbitals into low-lying $t_{2g}$ quenched states
and active $e_{g}$ states. The model of the undoped system includes
hopping $t$ along $\sigma$ bonds between $3z^{2}_{\gamma}-r^{2}$
orbitals, where $z_{\gamma}=x,y,z$ for the axis $\gamma=a,b,c$
\cite{Fei05}. The orthogonal $(x^{2}-y^{2})$-type orbitals do not
hybridize due to the symmetry of the underlying $p$-$d$ bonds. The
superexchange Hamiltonian between Cu ions is derived from
$d^{9}d^{9}\Leftrightarrow d^{8}d^{10}$ virtual charge excitations in
the presence of large $U$ \cite{Ole00}.
Therefore, KCuF$_{3}$ is described by a quintessential spin-orbital
model combining these two degrees of freedom in an essentially
isotropic 3D system. For vanishing Hund's exchange
it simplifies to two terms,
\begin{subequations}
  \label{eq:exch}
\begin{align}
    \label{eq:exch1}
H_{1}&=(J/S^2)\sum_{\mean{ij}\|\gamma}
    \left(\vect{S}_{i}\cdot\vect{S}_{j}+S^2\right)
    \left(\tau_{i}^{\gamma}\tau_{j}^{\gamma}+\tfrac{1}{4}\right),\\
    \label{eq:exch2}
H_{2}&=(J/2S^2)\sum_{\mean{ij}\|\gamma}
    \left(\vect{S}_{i}\cdot \vect{S}_{j}-S^2\right)
    \left(\tau_{i}^{\gamma}+\tau_{j}^{\gamma}\right),
\end{align}
\end{subequations}
where $J=t^2/U$, i.e., $J_{\rm Cu}=4t^2/U$ for $S=1/2$ spins.
The $T=1/2$ orbital operators depend on the direction $\gamma$:
$\tau^{a(b)}=-\frac{1}{2}(T^{z}\mp\sqrt{3}T^{x})$ and
$\tau^{c}=T^{z}$, under the standard convention \cite{vdB99} with
$\ket{3z^2-r^2}\equiv\ket{z}$, $\ket{x^2-y^2}\equiv\ket{\bar{z}}$.

Experimental data \cite{Bella,Pao02,Cac02,Dei08} as well as local spin
density approximation (LSDA) and LSDA+$U$ calculations
\cite{Bin04,Pav08,Leo10}, and the simulations of effective spin-orbital
model at large $U$ (with finite Hund's exchange $J_H$) find
\cite{Ole00} that the ground state of KCuF$_3$ has broken symmetry with
$A$-type AF ($A$-AF) and $C$-type AO ($C$-AO) order. Indeed, the energy
is gained from $H_1$ \eqref{eq:exch1} when either
$\langle \vect{S}_i\cdot\vect{S}_j\rangle>0$ and
$\langle\tau_i^{\gamma}\tau_j^{\gamma}\rangle<\frac14$, or
$\langle \vect{S}_i\cdot\vect{S}_j\rangle<S^2$ and
$\langle\tau_i^{\gamma}\tau_j^{\gamma}\rangle>0$,
as predicted by the Goodenough-Kanamori rules \cite{Goode,Kan59}. While
the latter occurs for AF bonds along the $c$ axis, the former stands for
FM spin order in $ab$ planes favored by finite Hund's exchange $J_H>0$.

An orbital crystal field $H_{z}=E_{z}\sum_{i}T_{i}^{z}$ is equivalent
to an axial pressure along the $c$ axis. It controls the AO order with
orbitals selected to minimize the energy \cite{Bal01}. For convenience
we take the classical ground state $\ket{0}$ for
$\mean{H_{2}+H_{z}}_{\mathrm{cl}}=0$ \footnote[2]{In experimental
systems AO order depends on $E_z$ \cite{Bal01}.},
implying that the orbitals $\{\ket{z},\ket{\bar{z}}\}$ are degenerate
and the occupied hole states in the AO state of KCuF$_3$ are
\cite{vdB99}: $\ket{\pm}=\frac{1}{\sqrt{2}}(\ket{z}\pm\ket{\bar{z}})$.

We now build the Hamiltonian for an electron doped into $\ket{0}$.
Let $d_{i,mn}^{\dag}$ be creation operators for an electron at site
${i}$. If $m=n=0$, the electron is added to the $d^{9}$ ground state
configuration, else it is added to the orbital-excited state if $n=1$
and/or to the spin-excited state if $m=1$. We decompose
$d_{i,mn}^{\dag}=f_{i}^{\dag}a_{i}^{n}b_{i}^{m}$, where $f_i^{\dag}$
is a fermion operator for the $d^{10}$ state, and $a_i$ and $b_i$ are
boson operators for orbital and spin excitations.
Following the same procedure, we find these additional terms when an
electron is doped in the system:
\begin{subequations}
  \label{eq:kin}
  \begin{align}
  \label{eq:T}
    \mathcal{T} &= -\frac{t}{4}\sum_{\langle ij \rangle\perp c}
    (f_{i}^{\dag}f_{j}^{} + \mathrm{H.c.}) = \sum_{\vect{k}} \epsilon_{\vect{k}}^{}
    f_{\vect{k}}^{\dag} f_{\vect{k}}^{},\\
    \label{eq:V1}
    \mathcal{V}_{\perp}^{t} &= -\frac{t}{4} \sum_{i,\delta \perp c}
    \left[
    (2+\sqrt{3} e^{i\pi_{y}\delta} e^{i \vect{Q} \vect{R}_{i}}) a_{i}^{\dag}
    + a_{i}^{\dag} a_{i+\delta}^{} +\right.\nonumber\\
     &\left.+(2-\sqrt{3} e^{i\pi_{y}\delta} e^{i\vect{Q} \vect{R}_{i}}) a_{i+\delta}^{}
    \right]
     (1+b_{i}^{}b_{i+\delta}^{\dag}) f_{i}^{\dag} f_{i+\delta}^{} +\mathrm{H.c.}\nonumber\\
  &-\frac{t}{4}
    \sum_{\mean{ij}\perp c} b_{i}^{}b_{j}^{\dag} f_{i}^{\dag} f_{j}^{}
    + \mathrm{H.c.},\\
     \label{eq:V2}
    \mathcal{V}_{\parallel}^{t} &=
    -\frac{t}{2}\!\sum_{\mean{ij}\parallel c}
    (a_{i}^{}-1)(a_{j}^{\dag}-1)(b_{i}^{}+b_{j}^{\dag})f_{i}^{\dag}f_{j}^{}
    +\mathrm{H.c.},
\end{align}
\end{subequations}
where $\epsilon_{\vect{k}}=-t(\cos k_x+\cos k_y)/2$, $\vect{Q}=(\pi,\pi,0)$,
and $\pi_{y}=(0,\pi,0)$, with the lattice constant $a=1$.

Free fermion propagation \eqref{eq:T} is allowed within the $ab$ planes
--- it does not modify the AO order \cite{vdB00}. Other processes which
contribute to in-plane fermion hopping are accompanied by creation or
removal of orbitons, as well as moving magnons around
(if any are present) \eqref{eq:V1}. In contrast, along the $c$ axis
free fermion hopping is blocked by the AF spin order, thus a magnon
(spin-flip excitation) always accompanies fermion hopping
between planes; orbitons may also be involved, see Eq. \eqref{eq:V2}.

\section{Variational approximation}
\label{sec:var}

We use a well-established
variational method \cite{Ber06,Mar10,Ber11,Ebr15} to determine the
one-electron Green's function
$G(\vect{k},\omega)=\bra{\vect{k}}\mathcal{G}(\omega)\ket{\vect{k}}$,
where $\mathcal{G}(\omega)=[\omega+i\eta-\mathcal{H}]^{-1}$ is the
resolvent operator and $\ket{\vect{k}}=f_{\vect{k}}^{\dag}\ket{0}$
is a free electron state doped into~$\ket{0}$.
The Hamiltonian is divided into
$\mathcal{H}_{0}=\mathcal{T}+\mathcal{H}_{J}^{z}$,
with the Ising part of the exchange~\eqref{eq:exch} (we have
verified that the quantum fluctuations are of little importance and
we ignore them, also see \cite{Bie16}), and the interaction
$\mathcal{V}= \mathcal{V}_{\perp}^{t}+ \mathcal{V}_{\parallel}^{t}$.

The variational method uses Dyson's equation
$\mathcal{G}(\omega)=\mathcal{G}_{0}(\omega)
+\mathcal{G}(\omega)\mathcal{V}\mathcal{G}_{0}(\omega)$ to generate
equations of motion (EOMs) for the Green's functions, within the
chosen variational space. Specifically, evaluation of
$\mathcal{V}\ket{\vect{k}}$ in real space links to generalized
propagators that have various bosons beside the fermion. The EOMs for
the corresponding generalized Green's functions are also obtained using
the Dyson's equation; the action of $\mathcal{V}$ links to new states
with increasingly more bosons. To close the set of EOMs, the hierarchy
is truncated by forbidding boson
configurations not included in the variational space.

This method generates analytical EOMs that implement the local
constraints exactly (\emph{e.g.}, not having the electron at a site
that hosts an orbiton or a magnon). Once generated, the EOMs can be
solved numerically to yield all the Green's functions, and in particular
$G(\vect{k},\omega)$ from which one can determine the spectral
function,
\begin{equation}
A(\vect{k},\omega)=-\frac{1}{\pi}\Im G(\vect{k},\omega),
\end{equation}
directly related to angle resolved photoemission spectroscopy
for LaMnO$_3$ or inverse photoemission for KCuF$_3$.
It needs to be emphasized, however, that the present model
employs a number of idealizations and was not intended to
produce a realistic low energy excitation spectrum, but rather
study the effects of spin and orbital excitations on the charge
dynamics in systems with the $A$-AF/$C$-AO ground state as
encountered in KCuF$_{3}$ and LaMnO$_{3}$. The present results
are therefore not addressing the experiment.
The accuracy of the results can be systematically improved by
increasing the variational space until convergence is achieved,
see also the Appendix.

\section{Results and discussion}
\label{sec:res}

\begin{figure}[t!]
  \centering
  \includegraphics[width=\columnwidth]{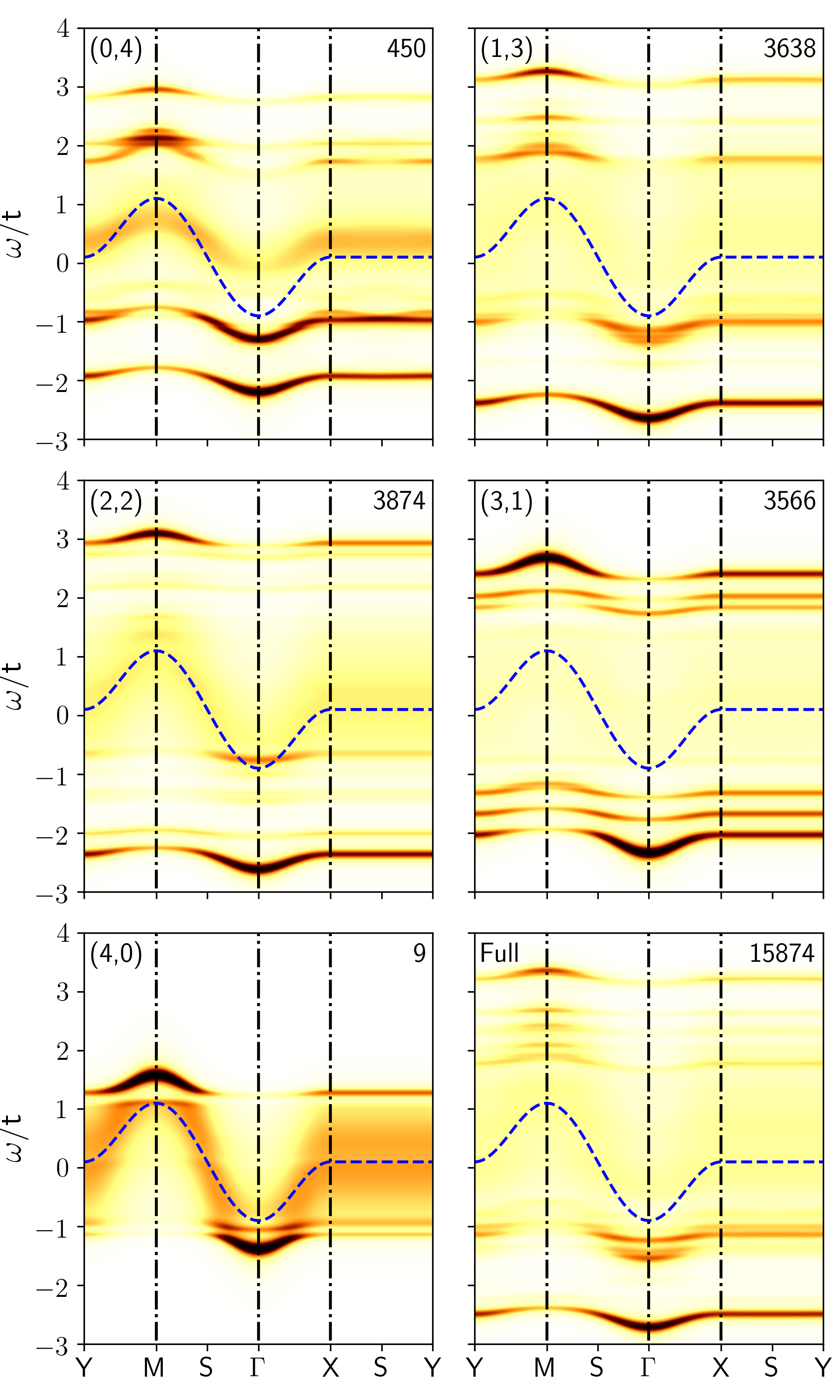}
  \caption{Spectral functions (3) for the case of the Ising model for
    $J=0.1$ and $S=\tfrac12$, presented in nonlinear $\tanh$-scale.
    The dashed blue line indicates free electron energy
$E_{\vect{k}}=\epsilon_{\vect{k}}+J$. The number in the upper-right
corner indicates the number of states generated in the EOM expansion.}
  \label{fig:ising1}
\end{figure}

Figure \ref{fig:ising1} shows
$A(\vect{k},\omega)$ contour plots for different
cutoffs $(m,n)$ where $m$ and $n$ are the maximum number of allowed
magnons and orbitons, respectively. The last panel shows
results with up to 4 bosons of either type. Plots are presented in
nonlinear $\tanh$-scale to highlight the low-weight incoherent part.

Clearly, the spectrum changes significantly between different
variational subspaces. In all cases there is a broad central continuum,
roughly overlapping the free particle bandwidth (dashed line). In the
$(0,4)$ orbiton-rich case there are ladders of discrete QP states
extending well below and above this continuum, consistent with the 2D
solution of the fermion-orbiton problem \cite{Bie16}. The magnon-rich
case $(4,0)$ has two QP bands closely sandwiching the continuum,
with a large transfer of spectral weight (see below) giving the
impression that QP pockets form only around $\Gamma$ and $M$.
The full case is a mix of both: it has one QP band below
and one above the continuum like the magnon-rich case, and both are
far from the continuum like in the orbiton-rich case.
The remaining ladder-states of the orbiton-rich case acquire finite
lifetimes and merge within a broader incoherent continuum.

\begin{figure}[t!]
  \centering \includegraphics[width=\columnwidth]{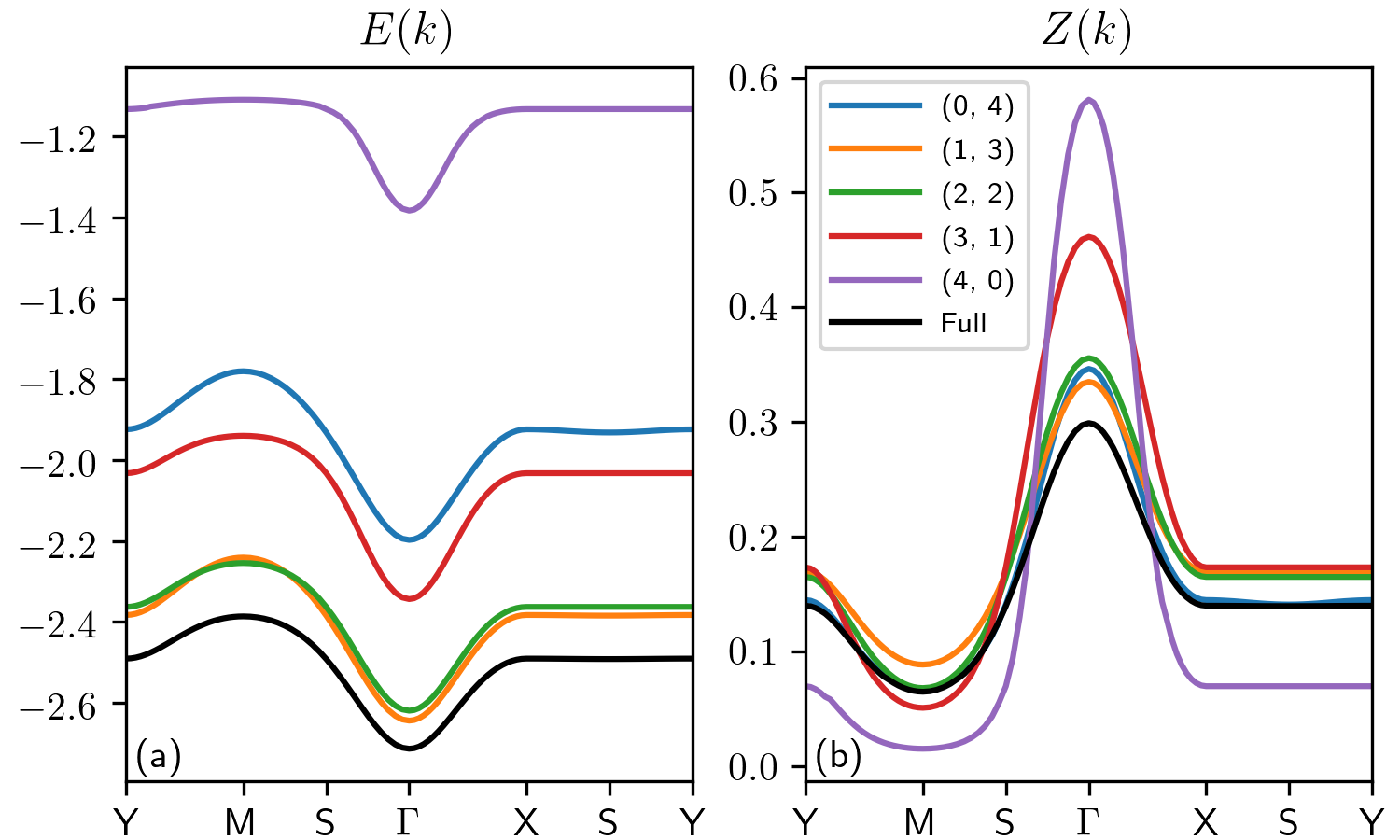}
\caption{Electron QP energy (left) and spectral weight of the 4th order
    solutions (right) for $J=0.1$ and $S=\frac12$.}
  \label{fig:conv}
\end{figure}

The shape and bandwidth of the low-energy QP band, however,
is remarkably unaffected by the structure of the cloud, as
shown in Fig.~\ref{fig:conv}(a). The magnon-only case has the
highest overall energy and the narrowest bandwidth due to being
pressed against the continuum. This explains the suppression of the
spectral weight at the M point, see Fig.~\ref{fig:conv}(b), and the
resulting impression of a $\Gamma$ QP pocket in
Fig.~\ref{fig:ising1}. The admixture of orbitons stabilizes the
polaron, pulling it to lower energies, but without affecting the band
shape. As expected, the full solution (the largest variational space)
has the lowest energy,  below that of the $(1,3)$ and $(2,2)$ subspaces.
This shows that there is
significant mixing between such configurations in the actual polaron
cloud. Thus, this is intrinsically a spin-orbital polaron that cannot
be fully described in terms of either spin-only or orbital-only models.

A major surprise comes when we consider
the evolution of the QP band upon addition of magnons to the cloud.
The zero-magnon case implies purely in-plane motion,
because magnons are only generated when the electron hops along the
$c$ axis. Indeed, here our results agree well with the 2D
orbital-polaron of Ref.~\cite{Bie16}. Naively, one would expect hopping
to another layer to make the polaron much heavier, if not outright
incoherent, because the magnon left behind is immobile in the Ising
limit. If this magnon is bound into the cloud then the
polaron would slow down significantly, whereas if it is unbound,
this should result in a finite QP lifetime as the polaron scatters off
of it. Indeed, Self-Consistent Born
Approximation results support this scenario \cite{Bal01}.

Our results clearly demonstrate that this naive expectation is wrong:
the 3D polaron is almost as mobile as its 2D counterpart. To
see why, we present results for much smaller variational spaces where
the EOMs are simple enough to allow analytical solutions that explain
the differences between magnons and orbitons.

\begin{figure}[t!]
  \centering \includegraphics[width=\columnwidth]{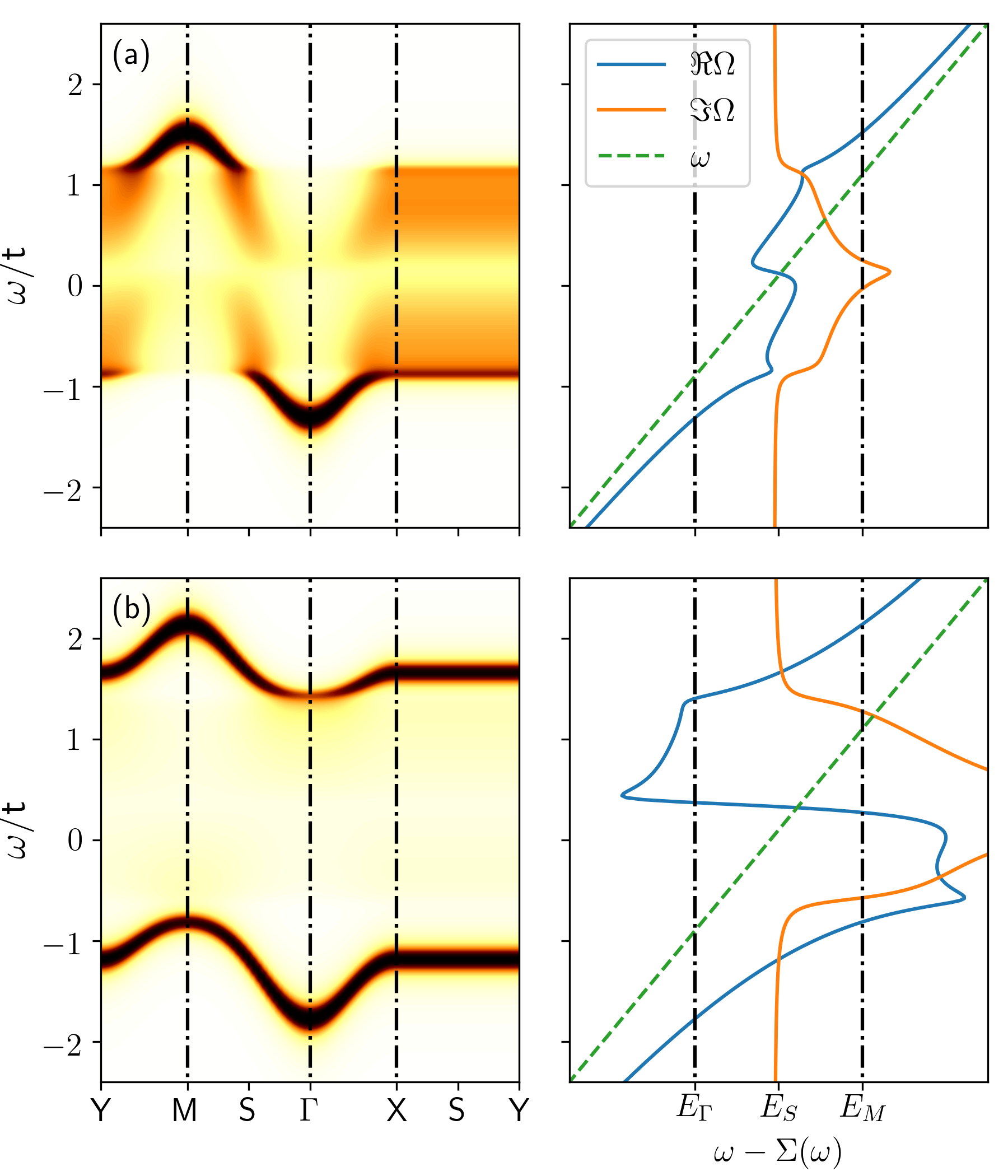}
\caption{Left--- Spectral functions (3) and
Right--- the real (blue) and imaginary (orange) parts of
$\omega - \Sigma(\omega)$ for variational result with a single:
(a) magnon and
(b) orbiton.
Parameters: $J=0.1$, $S=\frac12$, and $\eta=0.05$.
See text for more details.}
  \label{fig:sigma}
\end{figure}

Self-energies $\Sigma(\omega)$ for
(i)  one magnon and
(ii)  one orbiton allowed into the cloud are
derived in the Appendix.
They are linked to the propagator through
$G(\vect{k},\omega)=[\omega+E_{\vect{k}}-\Sigma(\vect{k},\omega)]^{-1}$,
where $E_{\vect{k}}=\epsilon_{\vect{k}}+J$ is the free electron
dispersion. Unlike for bigger variational spaces, in the $(0,1)$ and
$(1,0)$ cases the self-energy is momentum-independent making the
analysis particularly simple. The right panels of Fig.~\ref{fig:sigma}
display the real and imaginary parts of $\Omega=\omega-\Sigma(\omega)$
vs.~$\omega$. The vertical dash-dotted lines mark the upper/lower
bounds of $E_{\vect{k}}$. If $E_{\vect{k}} = \omega-\Sigma(\omega)$ for
an energy $\omega$ where $\mathrm{Im} \Sigma(\omega)=0$, then a QP with
infinite lifetime exists at this $(\vect{k},\omega)$ point.
If $\mathrm{Im} \Sigma(\omega)\ne 0$,
this is an incoherent state in the continuum.

The clear difference is that in the orbiton case (lower panel) there
is a QP solution well below the incoherent continuum for all $\vect{k}$,
whereas in the magnon case, there is a QP solution only for momenta
spanning the lower-half of the $E_{\vect{k}}$ range. The results are
thus qualitatively similar to those presented in Fig. \ref{fig:ising1}:
a spin-polaron squeezed just below the continuum as opposed to a
well-separated orbiton-polaron band. This is because $\Sigma(\omega)$
 is much smaller in the magnon than in the orbiton case
[$\mathrm{Re}\Sigma(\omega)$ is the difference between the blue line
showing $\omega - \mathrm{Re}\Sigma(\omega)$ and the dashed green line
showing $\omega$].

The one-magnon result is
$\Sigma_{(1,0)}(\omega)\!=\!
2(\frac{t}{2})^2G_{00}(\omega\!-\!\frac{3}{2}J)$. 
This can be understood as follows: the factor of 2 is because the
hopping along the $c$ axis can be either to the layer above or below;
$t/2$ is the effective hopping between layers, and is squared because
the electron must return to the original layer; finally,
$G_{nm}(\omega-\frac{3}{2}J)$ is the in-plane propagator for the free
electron to move a distance $(n,m)$ away from the site located just
above/below the magnon. Thus $G_{00}(\omega)$
controls how likely it is for the two objects to be adjacent.
In contrast, for the purely orbiton case,
$\Sigma_{(0,1)}(\omega)\,=\, 4(\frac{t}{2})^{2}[\,7G_{00}(\omega')
+2G_{11}(\omega') +7G_{20}(\omega')\,]/4$,
with $\omega'=\omega-4J$. There are now four in-plane hopping
directions and the hopping integrals are
$\frac{t}{2}(1\pm {\sqrt{3}\over 2})$,
hence the factors in front of propagators. Finally, there are more
ways for the orbiton to be absorbed, which is why several local
propagators appear. Below the free particle continuum these
propagators decrease exponentially with distance;
if we keep only the largest $G_{00}$ contribution, then here
$\Sigma_{(0,1)}(\omega) =3.5 \Sigma_{(1,0)}(\omega)$.
The difference is therefore due to an interplay between dimensionality
($ab$- vs. $c$-axis hopping) and the specific orbital order.

This also explains why adding magnons to the polaron cloud will not
change its dispersion substantially. Low-order self-energy diagrams
involving few bosons are generally non-crossed because
magnon and orbiton creation and absorption are accompanied by electron
motion in different spatial directions. Thus, to first order the
self-energy is the sum of the two separate contributions (higher order
crossed magnon-orbiton processes are possible but they involve several
bosons and therefore have a low probability and small contributions).
Clearly, adding the smaller magnon to the larger orbiton self-energy,
has only a limited effect on the QP band, pushing it to lower energies
but not changing its shape considerably.

We suggest that the present scenario might work
even better for LaMnO$_3$. The LSDA+$U$
calculations \cite{Leo10,Pav10} and model calculations
\cite{Fei99,Ish97,Oka02}
predict $A$-AF/$C$-AO order, as indeed observed
\cite{Kim03,Zhou,Kov10}. In this case the configuration
of Mn$^{3+}$ ions is $t_{2g}^{3}e_g^{1}$ and the leading superexchange
terms have the same $T=1/2$ pseudospins but $S=2$ spins~\cite{Fei99}
in Eqs.~\eqref{eq:exch} and thus spin
and orbital operators are almost disentangled~\cite{Sna16}. The bigger
spin $S=2$ is exactly counterbalanced by the reduced superexchange
$J_{\mathrm{Mn}}=J/4$; hence the absolute energy scale is roughly the
same. However, the spin-fermion coupling can only change $S=2$ to $S=1$
which means that the cost of magnons will be roughly 4 times smaller
than in KCuF$_3$, while the orbiton energy is simultaneously amplified
by the Jahn-Teller terms \cite{Fei98,Hot99,Sna16} making them
more classical. This implies that the hole spectral functions
for LaMnO$_3$ are similar to the electron ones for KCuF$_3$ but the
disproportion between magnons and orbitons is even stronger, with
less magnon coherence and amplified dominance of orbitons over magnons
for the mixed solutions. Indeed, it was found that orbital polarization
around a hole can lead to a very narrow QP band and to large incoherent
spectral weight \cite{Bal02}, indicating hole confinement
in a lightly doped LaMnO$_3$ insulator.

\section{Conclusions}
\label{sec:summa}

We presented variational solutions for the
polaron that forms in doped KCuF$_3$ or LaMnO$_3$. Comparison of
different cloud structures shows that the polaron has intrinsic
spin-orbital nature. The presence of magnons, however, is not
detrimental to the resulting QP speed and/or lifetime. This is a
direct consequence of the spin-orbital ground state that enforces the
creation of magnons when the electron (or hole) hops along the $c$
axis \cite{Mar91}. In contrast, the non-conservation of orbital flavor
allows for free in-plane electron (hole) propagation and for stronger
electron-orbiton interactions promoting robust polarons,
both in KCuF$_3$ and in LaMnO$_3$.

\begin{acknowledgments}
We thank Krzysztof Wohlfeld for insightful discussions. We kindly
acknowledge support by UBC Stewart Blusson Quantum Matter Institute,
by Natural Sciences and Engineering Research Council of Canada (NSERC),
and by Narodowe Centrum Nauki (NCN, National Science Centre, Poland)
under Projects No.~2012/04/A/ST3/00331 and 2015/16/T/ST3/00503.
\end{acknowledgments}

\appendix*

\section{Derivation of the one-boson solutions}
\label{sec:appendix}

We use a well-established variational method
\cite{Ber06,Mar10,Ber11,Ebr15} and allow only for a single excitation
by a propagating electron.
In general, generating the equations of motion (EOMs) in the
variational method proceeds along the same lines regardless of the
order of expansion. Therefore, this Appendix will also
serve to demonstrate the method itself to the unfamiliar reader.
The two solutions we have presented in the main text are the Green's
functions obtained with the generation of:\\
(a) a single magnon or \\
(b) a single orbiton.

We start from expanding the resolvent operator according to the Dyson
equation:
\begin{multline}
  \label{eq:Gk}
  \bra{\vect{k}}\mathcal{G}(\omega)\ket{\vect{k}}=
  \bra{\vect{k}}
      [1+\mathcal{G}(\omega)\mathcal{V}]\\
\times\frac{1}{\sqrt{N}}\sum_{i}e^{i\vect{k}\vect{R}_{i}} f_{i}^{\dag}\ket{0}
      G_{0}(\vect{k},\omega-J),
\end{multline}
where $G_{0}(\vect{k},\omega)=1/(\omega+i\eta-\epsilon_{\vect{k}})$ is
the free electron propagator. Next, we need to evaluate the result of
$\mathcal{V}f_{i}^{\dag}\ket{0}$, which will generate bosons in the
system.

In case (a) the boson stands for a magnon which results
from the fermion hopping along direction the $c$ axis according to
$\mathcal{V}_{\parallel}^{t}$. This leads to the following EOM for the
function $G(\vect{k},\omega)$:
\begin{equation}
  \label{eq:g1m}
  G(\vect{k},\omega) = \left[ 1-\frac{t}{2}\sum_{\delta\parallel c}
    F_{1}(\vect{k},\omega,\delta) \right] G_{0}(\vect{k},\omega-J),
\end{equation}
where
\begin{equation}
  \label{eq:f1}
  F_{1}(\vect{k},\omega,\delta) = \bra{\vect{k}} \mathcal{G}(\omega)
  \frac{1}{\sqrt{N}}\sum_{i} e^{i\vect{k}\vect{R}_{i}} b_{i}^{\dag}
  f_{i+\delta}^{\dag} \ket{0}
\end{equation}
is a generalized Green's function for a single magnon state. This
function is unknown and needs to be expanded as before, based on the
Dyson equation. However, this time $\mathcal{G}_{0}(\omega)$ acts on a
state with a magnon present, which means the electron cannot move far
away from its place, otherwise it cannot de-excite the system by
removing the magnon. This means that the system is no longer
translational invariant and the electron cannot propagate freely to
all sites in the system, but has to return back to the vicinity of the
magnon; thus $\vect{k}$ is no longer a good quantum number. Instead, the
electron has to be described in terms of real space Green's functions
\begin{equation}
  \label{eq:Gmn}
  G_{mn}(\omega) = \frac{1}{4\pi^{2}}\int\limits_{-\pi}^{\pi}d^{2}k
    G_{0}(\vect{k},\omega) e^{i\vect{k}\cdot(m,n,0)},
\end{equation}
which describe the propagation of an electron between two in-plane
sites, separated by a vector $(m,n)$ lying in the $ab$ plane.

It can be shown that they may be expressed in terms of complex
analytical continuation of elliptic integrals of the first and second
kind. In the case of a single magnon solution, the electron and the
magnon are in two neighboring planes, therefore the free electron
propagation will be described by the function $G_{(0,0)}(\omega)$.
This leads us to the following expansion:
\begin{equation}
  \label{eq:f1m}
  F_{1}(\vect{k},\omega,\delta) = -\frac{t}{2} \left[
    G(\vect{k},\omega)+F_{2}(\vect{k},\omega,\delta,\delta) \right]
  G_{00}(\omega-\tfrac{3}{2}J),
\end{equation}
where $F_{2}(\vect{k},\omega,\delta,\delta)$ is a generalized Green's
function for two magnons in a row. Normally, the expansion would also
include functions describing states involving orbitons, but we ignore
them in the purely magnonic solution in the lowest order. Moreover,
since we are investigating a single magnon solution, we set $F_{2}=0$,
thereby closing the EOM system, which now consists of three equations.
Solving this system of equations produces the result:
\begin{equation}
  \label{eq:Gmsol}
  G(\vect{k},\omega) = \left[ \omega+i\eta
  -\epsilon_{\vect{k}}-J-\tfrac{1}{2}t^{2}
    G_{00}(\omega-\tfrac{3}{2}J) \right]^{-1}.
\end{equation}

In case (b) of the one orbiton solution, the first expansion leads to:
\begin{multline}
  \label{eq:green}
    G(\vect{k},\omega) = \left[ 1 -\frac{t}{2}\sum_{\delta\perp c}
      F_{1}(\vect{k},\omega,\delta) \mp\frac{\sqrt{3}t}{4}\sum_{\delta\perp
        c} \bar{F}_{1}(\vect{k},\omega,\delta) \right]\\
    \times G_{0}(\vect{k},\omega-J).
\end{multline}
This time however, the $\delta$ summation goes over the four in-plane
directions. $F_{1}$ is defined similarly as before and there is
another function:
\begin{equation}
  \label{eq:f1}
  \bar{F}_{1}(\vect{k},\omega,\delta) = \bra{\vect{k}} \mathcal{G}(\omega)
  \frac{1}{\sqrt{N}}\sum_{i} e^{i(\vect{k}+\vect{Q})\vect{R}_{i}} b_{i}^{\dag}
  f_{i+\delta}^{\dag} \ket{0},
\end{equation}
where $\vect{Q}=(\pi,\pi,0)$ is the vector characterizing the orbital
order the ground state. Its explicit appearance in the equations is due
to the non-conservation of the orbital pseudospin.
Expansion of the $F_{1}$ functions yields:
\begin{multline}
  \label{eq:f1s}
  F_{1}(\vect{k},\omega,\delta) = -\frac{t}{4}\sum_{\gamma}
  \left[2G(\vect{k},\omega)\pm\sqrt{3}\bar{G}(\vect{k},\omega)\right]\\
  \times\tilde{G}_{0}(\vect{R}_{i}+\gamma,\vect{R}_{i}+\delta,\omega-4J),
\end{multline}
\begin{multline}
  \label{eq:f1bs}
  \bar{F}_{1}(\vect{k},\omega,\delta) = -\frac{t}{4}\sum_{\gamma}
  \left[2\bar{G}(\vect{k},\omega)\pm\sqrt{3}G(\vect{k},\omega)\right]\\
  \times\tilde{G}_{0}(\vect{R}_{i}+\gamma,\vect{R}_{i}+\delta,\omega-4J),
\end{multline}
where we have already neglected the higher order functions. Also, in
this analytical solution, the fermion-boson swap term $\propto
a_{i}^{}a_{j}^{\dag}f_{i}^{\dag} f_{j}^{}$ of
$\mathcal{V}_{\perp}^{t}$ is neglected for simplicity.

Furthermore,
this time the fermion and the boson are located in the same plane, so
the fermion can end up on any of the sites neighboring the boson, but
it is forbidden from entering the boson's site. Moreover, a careful
inspection of the Ising Hamiltonian shows that the cost of a state
with the fermion neighboring the boson is lower than that when the
fermion is located farther away. The real space Green's functions
have to be corrected for these effects, which is again done
using the Dyson equation, with the interaction made to cancel the
hopping elements to and from the boson's site:
\begin{equation}
  \label{eq:vi}
  \mathcal{V}_{i} = \frac{t}{4} \sum_{\epsilon}
  (f_{i}^{\dag} f_{i+\epsilon}^{} + \mathrm{H.c.})
  -\tfrac{3}{4}J\sum_{\epsilon} f_{i+\epsilon}^{\dag}f_{i+\epsilon}^{}.
\end{equation}

The Dyson expansion of the corrected Green's functions leads to the equation:
\begin{multline}
  \label{eq:gi}
  \tilde{G}_{0}(\vect{R}_{m},\vect{R}_{n},\omega) =
  G_{0}(\vect{R}_{m},\vect{R}_{n},\omega-\tfrac{9}{2}J)+\\
  +\sum_{\epsilon} \tilde{G}_{0}(\vect{R}_{m},\vect{R}_{i}+\epsilon,\omega)
    \left[\tfrac{t}{4}G_{0}(\vect{R}_{i},\vect{R}_{n},\omega-\tfrac{9}{2}J)\right.\\
 \left.-\tfrac{3}{4}JG_{0}(\vect{R}_{i}+\epsilon,\vect{R}_{n},\omega-\tfrac{9}{2}J)\right].
\end{multline}
Unlike the regular real space Green's functions $G_{mn}$ which are
defined by the propagation vector $(m,n)$, the corrected functions
$\tilde{G}_{0}$, depend on both the starting and final positions
explicitly, since the presence of the boson at site $\vect{R}_{i}$
breaks the translational symmetry of the system. More details on the
practical details of solving the above equation can be found in
\cite{Bie16}.

\begin{figure}[t!]
  \centering \includegraphics[width=\columnwidth]{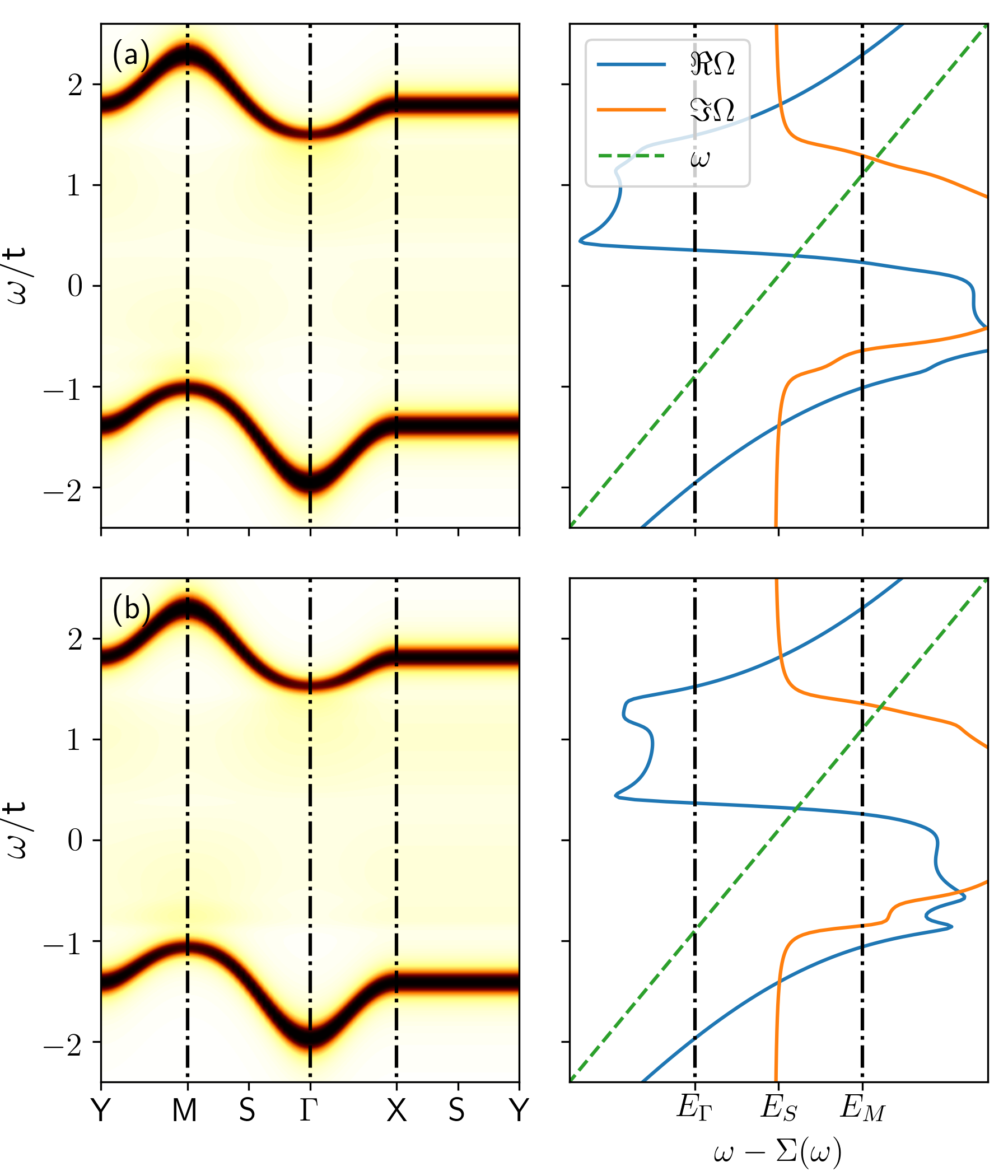}
\caption{Spectral function (left) and self-energy plots (right)
for the full 1st order solutions, without renormalization (a),
and with renormalization (b), resulting from the
$\mathcal{V}_{\parallel}^{t}$ Hamiltonian.
Parameters: $J=0.1$, $S=\frac12$.
}
  \label{fig:sigma2}
\end{figure}

Finally, the function
$\bar{G}(\vect{k},\omega)=
\bra{\vect{k}}\mathcal{G}(\omega)\ket{\vect{k}+\vect{Q}}$
expands into:
\begin{multline}
  \label{eq:gb}
  \bar{G}(\vect{k},\omega) = -\frac{t}{4}\sum_{\delta}
  \left[2\bar{F}_{1}(\vect{k},\omega,\delta)
    \pm\sqrt{3}F_{1}(\vect{k},\omega,\delta)\right]\\
  \times G_{0}(\vect{k}+\vect{Q},\omega-J).
\end{multline}
In total, this makes ten equations, which can be solved to yield,
after some simplification:
\begin{multline}
  \label{eq:Gmsol}
  G(\vect{k},\omega) = \left[
    \omega+i\eta-\epsilon_{\vect{k}}-J-\tfrac{1}{2}t^{2}G_{(1,1)}
    (\omega-4J)\right.\\
  \left.
    -\tfrac{7}{4}t^{2}(G_{(0,0)}(\omega-4J)+G_{(2,0)}(\omega-4J))
  \right]^{-1}.
\end{multline}

If we would now consider the full first order solution, \emph{i.e.}, up to one boson of any kind, we will find that in this case the self-energies are a simple sum of the single flavor solutions, since there are no processes linking the two sectors of the variational space. Furthermore, there will also be a renormalization of the magnonic self-energy resulting from the $a_{i+\delta}^{\dag}b_{i}^{}f_{i}^{\dag}f_{i+\delta}^{}$ process in the $\mathcal{V}_{\parallel}^{t}$ Hamiltonian, which produces a new state, with the fermion and the orbiton located in different planes:
\begin{multline}
  \label{eq:self2}
  \Sigma_{1}(\omega) = \tfrac{1}{2}t^{2}G_{(1,1)}(\omega-4J)\\
  +\tfrac{7}{4}t^{2}(G_{(0,0)}(\omega-4J)+G_{(2,0)}(\omega-4J))\\
  +\tfrac{1}{2}t^{2}G_{00}(\omega-\tfrac{3}{2}J)/
  [1-\tfrac{1}{4}t^{2}G_{00}(\omega-4J)].
\end{multline}

Figure~\ref{fig:sigma2} presents the spectral function and the self-energy plots for these full first order solutions. Clearly, the  mixing of magnons and orbitons does not lead to loss of coherence, and this solution is indeed qualitatively very similar to the purely orbitonic one, shown in  Fig. 3(b) in the main text.  This result supports the explanation given in  the main text for the rather small effect of the magnons on the  QP dispersion, as being due to the much smaller contribution of magnons to the self-energy because of different dimensionality and the specific ground-state orbital order.

\bibliographystyle{apsrev4-1}

%

\end{document}